\documentclass[12pt]{article} 
\usepackage{cite}
\usepackage{wrapfig}
\usepackage{amssymb}
\usepackage{amsfonts}
\usepackage{authblk}
\usepackage{amsmath}
\usepackage{lineno}
\usepackage{color}
\usepackage{xcolor}
\usepackage{multirow}
\usepackage{longtable}
\usepackage{lscape}
\usepackage[T2A]{fontenc}
\usepackage[utf8x]{inputenc}
\usepackage{hyperref}
\usepackage{bigstrut}

\usepackage{ifpdf}
\ifpdf
  \usepackage[pdftex]{graphicx}
  \usepackage{epstopdf}
\else
  \usepackage{graphicx}
\fi
\DeclareGraphicsExtensions{.png,.pdf,.jpg,.mps,.eps}
\graphicspath{{./figs/}}

\usepackage{floatrow}
\DeclareFloatFont{small}{\small}% "scriptsize" is defined by floatrow, "tiny" not
\usepackage[font=small,labelfont=rm]{caption}

\def\bqa {\begin{eqnarray}}
\def\eqa {\end{eqnarray}}
\newcommand{\dd}{\mathrm{d}}
\def\order#1{{\mathcal O}\left(#1\right)}
\def\Li#1#2{{\mathrm{Li}}_{#1}\left(#2\right)}
\usepackage{comment}

\begin{document}
\title{Electroweak effects in $e^+ e^- \to Z H$ process}

\author[1,2]{Andrej Arbuzov}
\author[1]{Serge Bondarenko}
\author[3]{Lidia Kalinovskaya}
\author[3]{Renat Sadykov}
\author[3,4]{Vitaly Yermolchyk}

\affil[1]{\small Bogoliubov Laboratory of Theoretical Physics, JINR,
                 141980 Dubna, Moscow region, Russia}
\affil[2]{\small Dubna State University,  
                 Universitetskaya str. 19,  141982 Dubna, Russia}
\affil[3]{\small Dzhelepov Laboratory of Nuclear Problems, JINR,  
                 141980 Dubna, Moscow region, Russia}
\affil[4]{\small Institute for Nuclear Problems, Belarusian State
                 University, Minsk, 220006 Belarus}

\date{}

\maketitle

% Author Orchid ID: enter ID or remove command
%\newcommand{\orcidauthorA}{0000-0001-9326-6905} % Add \orcidA{} behind the author's name
%\newcommand{\orcidauthorB}{0000-0001-7333-6275} % Add \orcidB{} behind the author's name
%\newcommand{\orcidauthorC}{0000-0001-5976-7933} % Add \orcidB{} behind the author's name
%\newcommand{\orcidauthorD}{0000-0002-9157-6819}
%\newcommand{\orcidauthorE}{0000-0002-0709-4228}

% Abstract (Do not insert blank lines, i.e. \\) 
\abstract{Electroweak radiative corrections to the cross section of the process $e^+ e^- \to Z H$ are considered. 
The complete one-loop electroweak radiative corrections are evaluated with the help of the SANC system.
Higher-order contributions of the initial state radiation are computed in the QED structure function formalism.
Numerical results are produced by a new version of the ReneSANCe event generator and MCSANCee integrator 
for the conditions of future electron-positron colliders. 
The resulting theoretical uncertainty in the description of this process is estimated. \\[.3cm]
}

\section{Introduction}

The Standard Model (SM) is extremely successful in describing
various phenomena in particle physics. In spite of this fact,
there are many reasons to consider the SM as an effective 
model, i.e., a low-energy approximation of a more general 
theory. Looking for the limits of the SM applicability domain is
one of the most valuable problems in modern fundamental physics.
On the other hand, a deep investigation of the SM properties
at the quantum level is still an important task since this model
is relevant for many applications in high-energy physics 
as well as astrophysics and cosmology. In this context, 
exploring the Higgs boson sector of the SM is crucial
for checking the mechanism of spontaneous symmetry breaking and
finalizing the verification of the model within the energy range
achieved at modern accelerators. 

To perform an in-depth verification of the Standard Model and 
define the energy region of its applicability, we certainly need 
a new high-energy accelerator. An electron-positron collider with energies
of a few hundred GeV looks now to be the best option. Several projects of
this kind of machine are being under consideration, e.g., 
ILC~\cite{Baer:2013cma}, 
CLIC~\cite{Linssen:2012hp},
CEPC~\cite{CEPCStudyGroup:2018ghi},
and FCC-ee~\cite{Abada:2019zxq}.
Programs of all these colliders necessarily include the option to run
as Higgs factories with center-of-mass system (c.m.s.) energies 
of about 240~GeV. At this energy, the maximal count rate of events of the
$e^+e^-\to ZH$ processes can be reached. Collecting several 
million such events will substantially increase the 
precision of the Higgs boson mass and the determination of the
partial decay widths~\cite{Fan:2014vta,An:2018dwb}.

The expected high statistics of events with Higgs bosons challenges
the theory to provide very accurate SM predictions for the corresponding
processes with uncertainty at the permille level. So we need to take
into account radiative corrections in the first and higher orders 
of perturbation theory\footnote{Non-perturbative effects due to strong interactions
are also relevant in running EW couplings and in producing extra meson pairs.}
The status of high-precision calculations for FCC-ee (and other
future $e^+e^-$ colliders in general) is described in~\cite{Blondel:2019vdq}.

In this work, we analyse QED and electroweak (EW) radiative corrections to
the higgsstrahlung process 
\begin{equation}
    e^+\ +\ e^-\ \to Z\ + H.
\label{HZprocess}
\end{equation}
This process is the most promising one in studying the Higgs boson
properties. So the accuracy of its theoretical description should be
higher than the experimental precision so that the combined uncertainty 
in the results of data analysis would not be spoiled by the theory.
The uncertainty estimate should be as complete as possible.

In this paper, we evaluate the complete one-loop corrections supplemented
by higher-order (HO) QED contributions in the leading logarithmic 
approximation (LLA)~\cite{Kuraev:1985hb}.
Our aim is to analyze the size of different HO contributions, estimate the resulting
theoretical uncertainty, and verify the necessity to include other HO
corrections. Note that in this work we do not consider decays of $Z$ and $H$, which 
are left for further study.

The complete one-loop electroweak radiative corrections to the process under
consideration were computed with the help of the SANC computer system and reported
in~\cite{Bondarenko:2018sgg}. Here we will concentrate on the analysis of the HO
QED effects. 

Two-loop QED corrections due to the initial state radiation (ISR) for a general process of 
high-energy electron-positron annihilation through a virtual photon or $Z$ boson were 
calculated in~\cite{Berends:1987ab} and recently corrected in~\cite{Blumlein:2020jrf}. 
Higher-order QED ISR contributions in the leading and next-to-leading 
logarithmic approximations up to the order $\mathcal{O}(\alpha^6L^5)$ were given 
in~\cite{Ablinger:2020qvo}. These results are performed in an inclusive set-up where 
only the distribution in the invariant mass of the final state particles is available. 
So they provide a benchmark for comparisons while for practical applications 
one needs a Monte Carlo simulation with complete kinematics.

The paper is organized as follows. In Section \ref{outline}, we outline the
contributions due to the higher-order QED initial state radiation order by order. 
In Section~\ref{NumRes}, we present the numerical results
for the cross section of associated $ZH$ production in the energy region
$\sqrt{s} = 200-500$ GeV.
Our conclusions are given in Section~\ref{conclusion}.

\section{ISR corrections in LLA approximation \label{outline}}

\subsection{General notes}

Let us consider ISR corrections to processes of high-energy
electron-positron annihilation within the LLA. They can be evaluated
with the help of the QED structure function formalism~\cite{Kuraev:1985hb}.
For ISR corrections in the annihilation channel the large 
logarithm is $L=\ln({s}/{m_e^2})$ where the total c.m.s.
energy $\sqrt{s}$ is chosen as factorization scale. 

The master formula reads
\begin{eqnarray} \label{master}
\sigma^{\mathrm{LLA}} =
\int\limits^{1}_{0}\dd x_1 \int\limits^{1}_{0}\dd x_2\ 
\mathcal{D}_{ee}(x_1) \mathcal{D}_{ee}(x_2)
\sigma_0(x_1,x_2,s) \Theta(\mathrm{cuts}),
\end{eqnarray}
where $\sigma_0(x_1,x_2,s)$ is the Born level cross section of the
annihilation process with reduced energies of the incoming
particles. 
Here we do not take into account ``photon induced''
contributions, since the corresponding kernel cross sections
$\sigma(\gamma e\to e ZH)$ and $\sigma(\gamma\gamma\to ZH)$
are very much suppressed by extra powers of the fine structure constant
$\alpha$.

The electron structure functions $\mathcal{D}_{ee}$ describe the density
of probability
to find an electron with an energy fraction $x$ in the initial
electron~\cite{Kuraev:1985hb,Skrzypek:1992vk,Arbuzov:1999cq}. 
In the LLA approximation we can separate the pure photonic corrections
(marked ``$\gamma$'') 
and the rest ones which include the pure pair and mixed photon-pair
effects (marked as ``pair'') as follows:
\begin{eqnarray} \label{DD}
&& \mathcal{D}_{ee}(x)= \mathcal{D}_{ee}^{\gamma}(x) 
+ \mathcal{D}_{ee}^{\mathrm{pair}}(x),
\\
&& \mathcal{D}_{ee}^{\gamma}(x)= \delta(1-x)
+ \frac{\alpha}{2\pi}(L-1)P^{(1)}(x)
+ \left(\frac{\alpha}{2\pi}(L-1)\right)^2\frac{1}{2!}P^{(2)}(x)
\nonumber \\ && \qquad
+ \left(\frac{\alpha}{2\pi}(L-1)\right)^3\frac{1}{3!}P^{(3)}(x)
+ \left(\frac{\alpha}{2\pi}(L-1)\right)^4\frac{1}{4!}P^{(4)}(x)
+ \order{\alpha^5L^5},
\\
&& \mathcal{D}_{ee}^{\mathrm{pair}}(x)= 
 \left(\frac{\alpha}{2\pi}L\right)^2\biggl[\frac{1}{3}P^{(1)}(x)
+ \frac{1}{2}R_s(x) \biggr]
\nonumber \\ && \qquad
+ \left(\frac{\alpha}{2\pi}L\right)^3\biggl[\frac{1}{3}P^{(2)}(x)
+ \frac{4}{27}P^{(1)}(x) 
+ \frac{1}{3}R_p(x) - \frac{1}{9}R_s(x) \biggr]
+ \order{\alpha^4L^4},
\end{eqnarray}
Pair corrections can be separated into singlet 
$(\sim R_{s,p})$ and non-singlet ones $(\sim P^{(n)})$.
We take into account both by default. 

Non-singlet splitting functions can be represented in the form
\begin{eqnarray} \label{PP}
P^{(n)}(x) = \lim_{\Delta\to 0}\biggl\{
\delta(1-x)P^{(n)}_\Delta + P^{(n)}_\Theta(x)\Theta(1-\Delta-x) \biggr\}
\end{eqnarray}
with $\Delta\ll 1$ being the soft-hard separator.
For example,
\begin{eqnarray} \label{P1}
P^{(1)}_\Delta = 2\ln\Delta + \frac{3}{2}, \qquad
P^{(1)}_\Theta(x) = \frac{1+x^2}{1-x}\, .
\end{eqnarray}
Higher-order non-singlet pure photonic splitting functions are
obtained by iterations of convolution
\begin{eqnarray} \label{Pn}
P^{(n+1)}(x) = \int\limits_0^1\dd x_1 \int\limits_0^1\dd x_2\ \delta(x-x_1x_2)
P^{(n)}(x_1)P^{(1)}(x_2), 
\end{eqnarray}
see the details in Ref.~\cite{Arbuzov:1999cq}.

The singlet splitting functions $R_s$ and $R_p$ are not singular at $x\to 1$, 
so they do not contain $\Delta$ parts.
Explicit expressions for all relevant splitting functions 
are given in Ref.~\cite{Arbuzov:1999cq,Arbuzov:2019hcg}.

The Born level partonic cross section $\sigma_0(x_1,x_2,s)$ is
known in the partonic c.m.s. as $\sigma_{\mathrm{Born}}(\hat{s})$,
where $\hat{s}=x_1 x_2 s$. The transition from the partonic c.m.s.
into the laboratory reference frame is required if $x_1 x_2 \neq 1$.

Let us classify contributions with different kinematics:
\begin{itemize}

\item[I.]{$(SV)_1\times (SV)_2$~~ The Born kinematics: additional 
contributions to Born+Soft+Virt.}

\item[II.]{$H_1\times(SV)_2$~~ One hard photon collinear to the first initial 
particle with possible soft and/or virtual (Soft+Virt) corrections to the second one.
Hereafter ``One hard photon'' means ``at least one hard photon in the same direction''. } 

\item[III.]{$(SV)_1\times H_2$~~ Soft+Virt to the second initial particle
and one hard photon along the first initial particle.}
  
\item[IV.]{$H_1\times H_2$~~ One hard photon along the first initial particle
and one along the second one. }
  
\end{itemize}

Separation of hard and soft photon emission is provided by the dimensionless
parameter $\Delta \ll 1$ with typical values $10^{-3}$, $10^{-4}$.
Under all integrals relevant (process dependent) cuts on 
the lower values of $x_1$ and $x_2$ values should be applied. 

Application of representation~(\ref{PP}) in structure 
functions~(\ref{DD}) and their substitution into the
master equation~(\ref{master}) gives the corrected
cross section in the LLA approximation. We expand the 
result in $\alpha$ and look at the second, third, and 
fourth order contributions. 

A few general comments are in order:

$\bullet$ The first lower index below denotes the order in $\alpha L$.  

$\bullet$ Factorials and coefficients are given explicitly in order
to see their structure.

$\bullet$ For pure photonic LLA corrections the traditional shift 
$L\longrightarrow (L-1)$ is carried out, it takes into account
part of the next-to-leading (NLO) corrections. However, for pair corrections
such a shift is not well justified and we keep the large log
unchanged.

\subsection{First order LLA contributions}

There are only photonic corrections in $\order{\alpha}$.
Below we list the contributions of different kinematics.

I. Born kinematics
\begin{eqnarray} 
&&\!\! \delta\sigma_{1,\gamma}^{(I)} =
\biggl(\frac{\alpha}{2\pi}(L-1)\biggr)\sigma_0(1,1,s)
\biggl\{ 2 P_\Delta^{(1)} \biggr\}.
\end{eqnarray} 

II. Emission only along the first particle
\begin{eqnarray} 
&&\!\!\delta\sigma_{1,\gamma}^{(II)} =
\biggl(\frac{\alpha}{2\pi}(L-1)\biggr)\!\!
\int\limits_{0}^{1-\Delta}\dd x_1 \sigma_0(x_1,1,s)
\biggl\{ P_\Theta^{(1)}(x_1) \biggr\}.
\end{eqnarray} 

III. Emission only along the second particle
\begin{eqnarray} 
&&\!\!\delta\sigma_{1,\gamma}^{(III)} =
\biggl(\frac{\alpha}{2\pi}(L-1)\biggr)\!\!
\int\limits_{0}^{1-\Delta}\!\!\dd x_2 \sigma_0(1,x_2,s)
\biggl\{ P_\Theta^{(1)}(x_2) \biggr\}.
\end{eqnarray} 

IV. Emission along both initial particles
\begin{eqnarray} 
\delta\sigma_{1,\gamma}^{(IV)} = 0.
\end{eqnarray}

\subsection{Second order LLA contributions}

I. Born kinematics
\begin{eqnarray} 
&&\!\! \delta\sigma_{2,\gamma}^{(I)} =
\biggl(\frac{\alpha}{2\pi}(L-1)\biggr)^2\sigma_0(1,1,s)
\biggl\{ 2\frac{1}{2!}P_\Delta^{(2)} + P_\Delta^{(1)}P_\Delta^{(1)} \biggr\},
\\
&&\!\!\delta\sigma_{2,\mathrm{pair}}^{(I)} =
\biggl(\frac{\alpha}{2\pi}L\biggr)^2\sigma_0(1,1,s)
\biggl\{ 2\frac{1}{3}P_\Delta^{(1)} \biggr\}.
\end{eqnarray} 

II. Emission only along the first particle
\begin{eqnarray} 
&&\!\!\delta\sigma_{2,\gamma}^{(II)} =
\biggl(\frac{\alpha}{2\pi}(L-1)\biggr)^2\!\!
\int\limits_{0}^{1-\Delta}\dd x_1 \sigma_0(x_1,1,s)
\biggl\{ \frac{1}{2!}P_\Theta^{(2)}(x_1) + P_\Theta^{(1)}(x_1)P_\Delta^{(1)} \biggr\},
\\
&&\!\!\delta\sigma_{2,\mathrm{pair}}^{(II)} =
\biggl(\frac{\alpha}{2\pi}L\biggr)^2\!\!
\int\limits_{0}^{1-\Delta}\!\!\dd x_1 \sigma_0(x_1,1,s)
\biggl\{ \frac{1}{3}P_\Theta^{(1)}(x_1) + \frac{1}{2}R_s(x_1) \biggr\}. 
\end{eqnarray} 

III. Emission only along the second particle
\begin{eqnarray} 
&&\!\!\delta\sigma_{2,\gamma}^{(III)} =
\biggl(\frac{\alpha}{2\pi}(L-1)\biggr)^2\!\!
\int\limits_{0}^{1-\Delta}\!\!\dd x_2 \sigma_0(1,x_2,s)
\biggl\{ \frac{1}{2!}P_\Theta^{(2)}(x_2) + P_\Theta^{(1)}(x_2)P_\Delta^{(1)} \biggr\},
\\
&&\!\!\delta\sigma_{2,\mathrm{pair}}^{(III)} =
\biggl(\frac{\alpha}{2\pi}L\biggr)^2\!\!
\int\limits_{0}^{1-\Delta}\!\!\dd x_2 \sigma_0(1,x_2,s)
\biggl\{ \frac{1}{3}P_\Theta^{(1)}(x_2) + \frac{1}{2}R_s(x_2) \biggr\}. 
\end{eqnarray} 

IV. Emission along both initial particles
\begin{eqnarray} 
&&\!\!\delta\sigma_{2,\gamma}^{(IV)} =
\biggl(\frac{\alpha}{2\pi}(L-1)\biggr)^2\!\!
\int\limits_{0}^{1-\Delta}\!\!\dd x_1 \!\!\!
\int\limits_{0}^{1-\Delta}\!\!\dd x_2 \sigma_0(x_1,x_2,s)
\biggl\{ P_\Theta^{(1)}(x_1)P_\Theta^{(1)}(x_2) \biggr\},
\\
&&\!\!\delta\sigma_{2,\mathrm{pair}}^{(IV)} = 0.
\end{eqnarray}

\subsection{Third order LLA contributions}

I. Born kinematics
\begin{eqnarray} 
&&\!\! \delta\sigma_{3,\gamma}^{(I)} =
\biggl(\frac{\alpha}{2\pi}(L-1)\biggr)^3\sigma_0(1,1,s)
\biggl\{ 2\frac{1}{3!}P_\Delta^{(3)} + 2P_\Delta^{(1)}\frac{1}{2!}P_\Delta^{(2)} \biggr\},
\\
&&\!\!\delta\sigma_{3,\mathrm{pair}}^{(I)} =
\biggl(\frac{\alpha}{2\pi}L\biggr)^3\sigma_0(1,1,s)
\biggl\{ 2\frac{1}{3}P_\Delta^{(2)} 
+ 2\frac{4}{27}P_\Delta^{(1)}  
%\nonumber \\ && \qquad
+ 2\frac{1}{3}P_\Delta^{(1)} P_\Delta^{(1)} 
\biggr\}.
\end{eqnarray} 

II. Emission only along the first particle
\begin{eqnarray} 
&&\!\!\delta\sigma_{3,\gamma}^{(II)} =
\biggl(\frac{\alpha}{2\pi}(L-1)\biggr)^3\!\!
\int\limits_{0}^{1-\Delta}\!\!\dd x_1 \sigma_0(x_1,1,s)
\biggl\{ \frac{1}{3!}P_\Theta^{(3)}(x_1) 
+ \frac{1}{2!}P_\Theta^{(2)}(x_1)P_\Delta^{(1)} 
\nonumber \\ && \qquad
+ P_\Theta^{(1)}(x_1)\frac{1}{2!}P_\Delta^{(2)} \biggr\},
\\
&&\!\!\delta\sigma_{3,\mathrm{pair}}^{(II)} =
\biggl(\frac{\alpha}{2\pi}L\biggr)^3\!\!
\int\limits_{0}^{1-\Delta}\!\!\dd x_1 \sigma_0(x_1,1,s)
\biggl\{ \frac{1}{3}P_\Theta^{(2)}(x_1) 
+ \frac{4}{27}P_\Theta^{(1)}(x_1) + \frac{1}{3}R_p(x_1) 
\nonumber \\ && \qquad
- \frac{1}{9}R_s(x_1) 
+ 2\frac{1}{3}P_\Theta^{(1)}(x_1) P_\Delta^{(1)} 
+ \frac{1}{2}R_s(x_1) P_\Delta^{(1)} 
\biggr\}. 
\end{eqnarray}

III. Emission only along the second particle
\begin{eqnarray} 
&&\!\!\delta\sigma_{3,\gamma}^{(II)} =
\biggl(\frac{\alpha}{2\pi}(L-1)\biggr)^3\!\!
\int\limits_{0}^{1-\Delta}\!\!\dd x_2 \sigma_0(1,x_2,s)
\biggl\{ \frac{1}{3!}P_\Theta^{(3)}(x_2) 
+ \frac{1}{2!}P_\Theta^{(2)}(x_2)P_\Delta^{(1)} 
\nonumber \\ && \qquad
+ P_\Theta^{(1)}(x_2)\frac{1}{2!}P_\Delta^{(2)} \biggr\},
\\
&&\!\!\delta\sigma_{3,\mathrm{pair}}^{(II)} =
\biggl(\frac{\alpha}{2\pi}L\biggr)^3\!\!
\int\limits_{0}^{1-\Delta}\!\!\dd x_2 \sigma_0(1,x_2,s)
\biggl\{ \frac{1}{3}P_\Theta^{(2)}(x_2) 
+ \frac{4}{27}P_\Theta^{(1)}(x_2) + \frac{1}{3}R_p(x_2) 
\nonumber \\ && \qquad
- \frac{1}{9}R_s(x_2)   
+ 2\frac{1}{3}P_\Theta^{(1)}(x_2) P_\Delta^{(1)} 
+ \frac{1}{2}R_s(x_2) P_\Delta^{(1)} 
\biggr\}. 
\end{eqnarray} 

IV. Emission along both initial particles

\begin{eqnarray} 
&&\!\!\delta\sigma_{3,\gamma}^{(IV)} =
\biggl(\frac{\alpha}{2\pi}(L-1)\biggr)^3\!\!
\int\limits_{0}^{1-\Delta}\!\!\dd x_1 \!\!\!
\int\limits_{0}^{1-\Delta}\!\!\dd x_2 \sigma_0(x_1,x_2,s)
\biggl\{ \frac{1}{2!}P_\Theta^{(2)}(x_1)P_\Theta^{(1)}(x_2) 
\nonumber \\ && \qquad
+ P_\Theta^{(1)}(x_1)\frac{1}{2!}P_\Theta^{(2)}(x_2) 
\biggr\},
\\
&&\!\!\delta\sigma_{3,\mathrm{pair}}^{(IV)} = 
\biggl(\frac{\alpha}{2\pi}L\biggr)^3\!\!
\int\limits_{0}^{1-\Delta}\!\!\dd x_1 \!\!\!
\int\limits_{0}^{1-\Delta}\!\!\dd x_2 \sigma_0(x_1,x_2,s)
\biggl\{ \frac{1}{3}P_\Theta^{(1)}(x_1)P_\Theta^{(1)}(x_2) 
+ \frac{1}{2}R_s(x_1)P_\Theta^{(1)}(x_2) 
\nonumber \\ && \qquad
+ P_\Theta^{(1)}(x_1)\frac{1}{3}P_\Theta^{(1)}(x_2) 
+ P_\Theta^{(1)}(x_1)\frac{1}{2}R_s(x_2) 
\biggr\}.
\end{eqnarray}

\subsection{Fourth order LLA contributions}

Here we list only pure photonic contributions due to the smallness of 
pair corrections in the fourth order.

I. Born kinematics
\begin{eqnarray} 
&&\!\! \delta\sigma_{4,\gamma}^{(I)} =
\biggl(\frac{\alpha}{2\pi}(L-1)\biggr)^4\sigma_0(1,1,s)
\biggl\{ 2\frac{1}{4!}P_\Delta^{(4)} + 2P_\Delta^{(1)}\frac{1}{3!}P_\Delta^{(3)}
+ \left(\frac{1}{2!}P_\Delta^{(2)}\right)^2 \biggr\}.
\end{eqnarray} 

II. Emission only along the first particle
\begin{eqnarray} 
&&\!\!\delta\sigma_{4,\gamma}^{(II)} =
\biggl(\frac{\alpha}{2\pi}(L-1)\biggr)^4\!\!
\int\limits_{0}^{1-\Delta}\!\!\dd x_1 \sigma_0(x_1,1,s)
\biggl\{ \frac{1}{4!}P_\Theta^{(4)}(x_1) 
+ \frac{1}{3!}P_\Theta^{(3)}(x_1)P_\Delta^{(1)} 
\nonumber \\ && \qquad
+ \frac{1}{2!}P_\Theta^{(2)}(x_1)\frac{1}{2!}P_\Delta^{(2)} 
+ P_\Theta^{(1)}(x_1)\frac{1}{3!}P_\Delta^{(3)} \biggr\}. 
\end{eqnarray} 

III. Emission only along the second particle
\begin{eqnarray} 
&&\!\!\delta\sigma_{4,\gamma}^{(II)} =
\biggl(\frac{\alpha}{2\pi}(L-1)\biggr)^3\!\!
\int\limits_{0}^{1-\Delta}\!\!\dd x_2 \sigma_0(1,x_2,s)
\biggl\{ \frac{1}{4!}P_\Theta^{(4)}(x_2) 
+ \frac{1}{3!}P_\Theta^{(3)}(x_2)P_\Delta^{(1)} 
\nonumber \\ && \qquad
+ \frac{1}{2!}P_\Theta^{(2)}(x_2)\frac{1}{2!}P_\Delta^{(2)} 
+ P_\Theta^{(1)}(x_2)\frac{1}{3!}P_\Delta^{(3)} \biggr\}. 
\end{eqnarray}

IV. Emission along both initial particles
\begin{eqnarray} 
&&\!\!\delta\sigma_{4,\gamma}^{(IV)} =
\biggl(\frac{\alpha}{2\pi}(L-1)\biggr)^3\!\!
\int\limits_{0}^{1-\Delta}\!\!\dd x_1 \!\!\!
\int\limits_{0}^{1-\Delta}\!\!\dd x_2 \sigma_0(x_1,x_2,s)
\biggl\{ \frac{1}{2!}P_\Theta^{(2)}(x_1)\frac{1}{2!}P_\Theta^{(2)}(x_2) 
\nonumber \\ && \qquad
+ P_\Theta^{(1)}(x_1)\frac{1}{3!}P_\Theta^{(3)}(x_2) 
+ \frac{1}{3!}P_\Theta^{(3)}(x_1)P_\Theta^{(1)}(x_2)
\biggr\}.
\end{eqnarray}

\subsection{LLA contributions for helicity states}

The leading order (LO) splitting function $P_{ee}(x)$ given
by Eq.~(\ref{P1}) preserves helicity~\cite{Altarelli:1977zs}, 
i.e.,
\begin{eqnarray} 
P^{(n)}_{e_+e_+}(x)=P^{(n)}_{e_-e_-}(x)=P^{(n)}_{ee}(x),
\qquad
P^{(n)}_{e_+e_-}(x)=P^{(n)}_{e_-e_+}(x)=0.
\end{eqnarray}

However, singlet contributions of pair corrections can
be separated for different helicities:
\begin{eqnarray} 
&& R_{s}(x)=R_{s,e_+e_+}(x)+R_{s,e_-e_+}(x),
\qquad
R_{p}(x)=R_{p,e_+e_+}(x)+R_{p,e_-e_+}(x),
\\ 
&& R_{s,e_+e_+}(x) = 3(1-x) + 2(1+x)\ln{x} + \frac{2(1-x^3)}{3x},
\\
&& R_{s,e_-e_+}(x) = - 2(1-x) + \frac{2(1-x^3)}{3x},
\\
&& R_{p,e_+e_+}(x) = 
- \frac{31}{6}(1-x)  
+ 6(1-x)\ln(1-x) 
- 2\ln{x} 
+ 4x\ln{x} 
+ \frac{4(1-x^3)}{3x}\ln(1-x)
\nonumber \\ && \qquad
+ \frac{4}{3}x^2\ln{x}
+ 4(1+x)\ln{x}\ln(1-x) 
- (1+x)\ln^2{x} 
+ 4(1+x)\Li{2}{1-x} ,
\\
&& R_{p,e_-e_+}(x) = 
\frac{4}{3}(1-x)
+ \frac{4(1-x^3)}{3x}\ln(1-x) 
- 4(1-x)\ln(1-x) 
\nonumber \\ && \qquad
+ 2(1-x)\ln{x} 
+ \frac{4}{3}x^2\ln{x}.
\end{eqnarray}
Since QED preserves parity,
\begin{eqnarray} 
R_{i,e_-e_+}(x) = R_{i,e_+e_-}(x), \qquad R_{i,e_-e_-}(x) = R_{i,e_+e_+}(x).
\end{eqnarray}

\subsection{Scheme with exponentiation}

In the master equation~(\ref{master}), the electron
structure functions can be taken in the exponentiated 
form~\cite{Cacciari:1992pz,Przybycien:1992qe}. That would mean continuous
integration over $x_{1,2}$ without the auxiliary parameter $\Delta$.
The result of~\cite{Przybycien:1992qe} contains only the pure photonic corrections and 
corresponds to the inclusion of exact
leading logs up to the order $\order{\alpha^5L^5}$ together with approximate
(incomplete) HO LLA effects. Note that the HO
exponentiated effects become exact (in LLA) in the soft photon limit.

$\bullet$ 
The pair LLA corrections can be added to the result of~\cite{Przybycien:1992qe} 
as, e.g., in Ref.~\cite{Arbuzov:1997pj} with a possible update for higher-order pair corrections listed above.

$\bullet$ 
The exponentiated structure functions include the LLA
part of one-loop QED radiative corrections. 
To avoid double counting with complete one-loop corrections,
we need to subtract the first order leading logarithmic terms 
from one-loop corrections. So the final result reads
\begin{eqnarray} \label{master_exp}
&& \sigma^{\mathrm{corr.}} =
\int\limits^{1}_{0}\dd x_1 \int\limits^{1}_{0}\dd x_2
\mathcal{D}_{ee}^{\mathrm{exp}}(x_1) \mathcal{D}_{ee}^{\mathrm{exp}}(x_2)
\sigma_0(x_1,x_2,s) \Theta(\mathrm{cuts})
\nonumber \\ && \qquad
+ \biggl[ \sigma^{\mathrm{Soft+Virt}} - \sigma_{\mathrm{LLA}}^{\mathrm{Soft+Virt}}\biggr]
+ \biggl[ \sigma^{\mathrm{Hard}} - \sigma_{\mathrm{LLA}}^{\mathrm{Hard}}\biggr].
\end{eqnarray}

The ``Soft+Virt'' part has the Born-like kinematics:
\begin{eqnarray} \label{SV_LLA}
\sigma_{\mathrm{LLA}}^{\mathrm{Soft+Virt}} = \sigma^{\mathrm{Born}}
2\frac{\alpha}{2\pi}(L-1)\left[2\ln\omega + \frac{3}{2}\right],  
\end{eqnarray}
where $\omega$ is the dimensionless soft-hard separator from the
original complete one-loop formulae implemented in the Monte-Carlo generator.

The ``Hard LLA'' term is rewritten to match "Hard" kinematics:
\begin{eqnarray} \label{H_LLA}
\sigma_{\mathrm{LLA}}^{\mathrm{Hard}} = \frac{\alpha}{2\pi}
\int \frac{\dd^3k}{k_0^2 2\pi}
\biggl[\frac{E^2}{kp_1}\sigma^{\mathrm{Born}}(x_1,1,s)
+ \frac{E^2}{kp_2}\sigma^{\mathrm{Born}}(1,x_2,s) \biggr]
\biggl(2-2\frac{k_0}{E}+\frac{k_0^2}{E^2}\biggr).
\end{eqnarray}

\section{Numerical results
\label{NumRes}}

In this section, we show numerical results for one-loop EW and HO
QED radiative corrections to the $e^+e^- \to ZH $ process. 
The input parameters can be found in~\cite{Bondarenko:2018sgg}.
%We use the following set of input parameters:
%\begin{eqnarray}
%\alpha^{-1}(0) &=& 137.03599976,
%\\
%M_W &=& 80.45150 \; \mathrm{GeV}, \quad M_Z = 91.1867 \; %\mathrm{GeV},
%\nonumber\\
%\Gamma_Z &=& 2.49977 \; \mathrm{GeV}, \quad m_e = 0.51099907 %\; \mathrm{MeV},
%\nonumber\\
%m_\mu &=& 0.105658389 \; \mathrm{GeV}, \quad m_\tau = 1.77705 %\; \mathrm{GeV},
%\nonumber\\
%m_d &=& 0.083 \; \mathrm{GeV}, \quad m_s = 0.215 \; %\mathrm{GeV},
%\nonumber\\
%m_b &=& 4.7 \; \mathrm{GeV}, \quad m_u = 0.062 \; %\mathrm{GeV},
%\nonumber\\
%m_c &=&
%1.5 \; \mathrm{GeV}, \quad m_t = 173.8 \; \mathrm{GeV}.
%\nonumber
%\end{eqnarray}
The results are obtained without any angular cuts.
The relative correction $\delta$ (in \%) is defined as
\begin{equation}
\delta = \frac{\sigma_{\tt LLA}}{\sigma^{\text{Born}}} - 1.
\end{equation}
    
\begin{figure}[ht]
\label{deltaL2L3L4LS}
\centering
\includegraphics[width=0.7\linewidth]{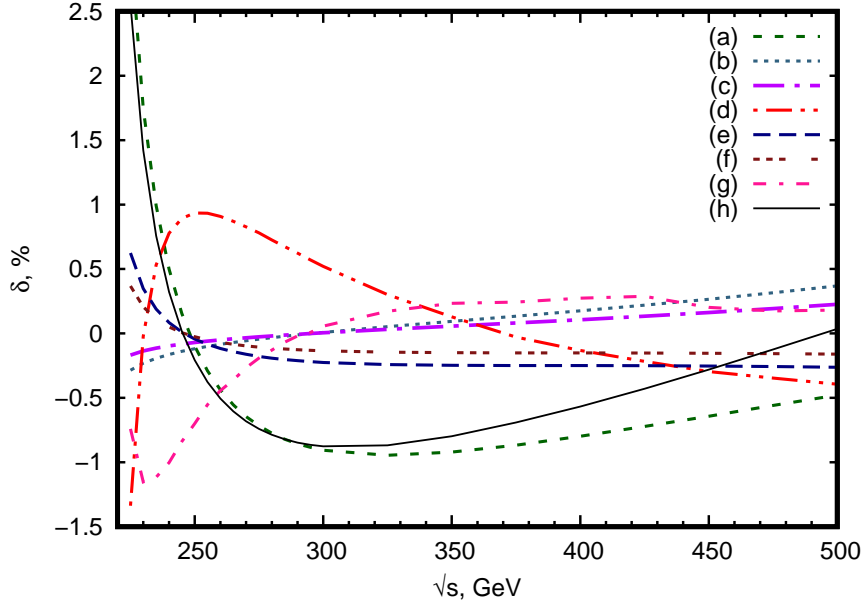}
\caption{\label{fig-deltaL2L3L4LS}
  Relative corrections (in \%) 
  (a) $\mathcal{O}({\alpha}^2L^2) {\gamma}$, 
  (b) $\mathcal{O}({\alpha}^2L^2) {e^+e^-}$,
  (c) $\mathcal{O}({\alpha^2}L^2) {\mu^+\mu^-}$,
  (d) 10 $\times$ $\mathcal{O}({\alpha}^3L^3) {\gamma}$, 
  (e) 10 $\times$ $\mathcal{O}({\alpha}^3L^3) {e^+e^-}$,
  (f) 10 $\times$ $\mathcal{O}({\alpha}^3L^3) {\mu^+\mu^-}$,
  (g) 100 $\times$ $\mathcal{O}({\alpha}^4L^4) {\gamma}$ 
  and (h) their sum vs. c.m.s. energy.
}
\end{figure}

To illustrate the trends of the ISR contribution behaviour, we present 
separate distributions for each $\mathcal{O}({\alpha}^nL^n)$, 
$n=2-4$ term and their sum as a function of the c.m.s. energy.
Figure~\ref{fig-deltaL2L3L4LS} shows the values of the contribution
of each relative correction (in \%) :
  (a) $\mathcal{O}({\alpha}^2L^2) {\gamma}$, 
  (b) $\mathcal{O}({\alpha}^2L^2) {e^+e^-}$,
  (c) $\mathcal{O}({\alpha^2}L^2) {\mu^+\mu^-}$,
  (d) 10 $\times$ $\mathcal{O}({\alpha}^3L^3) {\gamma}$, 
  (e) 10 $\times$ $\mathcal{O}({\alpha}^3L^3) {e^+e^-}$,
  (f) 10 $\times$ $\mathcal{O}({\alpha}^3L^3) {\mu^+\mu^-}$,
  (g) 100 $\times$ $\mathcal{O}({\alpha}^4L^4) {\gamma}$,
  and (h) their sum vs. c.m.s. energy.

The dominant contribution is $\mathcal{O}({\alpha}^2L^2) {\gamma}$ which is 
about 3\% at the c.m.s. energy $\sqrt{s}=220$ GeV, then it crosses the zero line 
approximately at $\sqrt{s}=250$ GeV and goes to $-0.5\%$ at $\sqrt{s}=500$ GeV. 
The second order contributions due to light pair emission are smaller and have 
the opposite sign. The third and fourth order corrections are approximately 
10 (100) times smaller, respectively. The sum is mainly determined by the
$\mathcal{O}({\alpha}^2L^2) {\gamma}$ 
term in the region of c.m.s energies $\sqrt{s}=220-290$ GeV
and becomes close to zero at $\sqrt{s}=500$ GeV.
One can see that in the threshold energy region there are several competing 
contributions with different behavior. This confirms the necessity to 
take into account HO QED ISR contributions in the studies of the
higgsstrahlung process at future colliders.

It is natural to consider the contributions of the corrections in some sets, that is, 
to compare the contributions of the same order of magnitude: (a)-(c), (d)-(f). 
The most significant value comes from  the photonic contributions (a) and (d).
The contributions from a pair is much less, but there is a kinematical region where 
their contribution must be taken into account. The suppression of pair corrections 
with respect to photonic ones in the same order is typical for high-energy annihilation
processes~\cite{Arbuzov:2001rt}.
  
\begin{figure}[ht]
\centering
\includegraphics[width=0.7\linewidth]{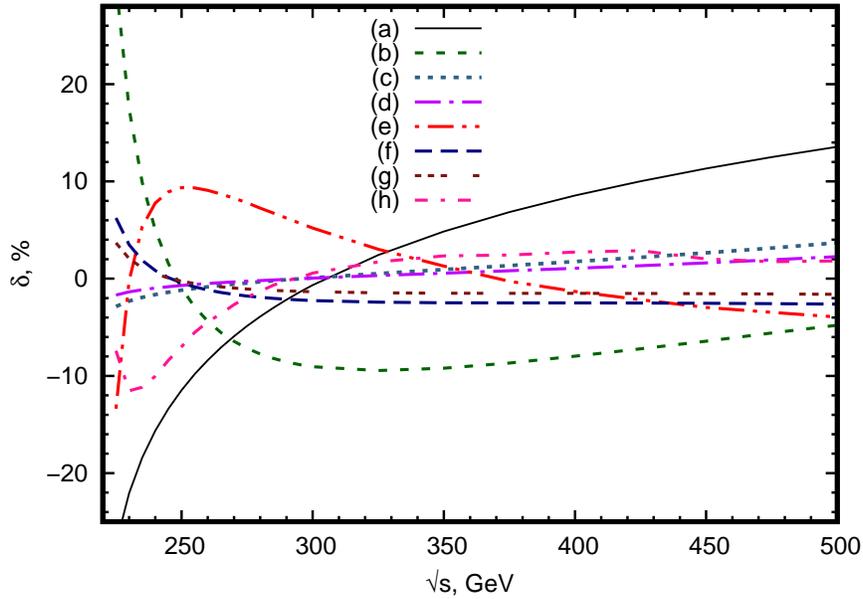}
\caption{\label{fig-delta-l1l2l3}
  Relative corrections (in \%) for the 
  (a) $\mathcal{O}({\alpha}L)$, 
  (b) 10 $\times$ $\mathcal{O}({\alpha}^2L^2) {\gamma}$, 
  (c) 10 $\times$ $\mathcal{O}({\alpha}^2L^2) {e^+e^-}$,
  (d) 10 $\times$ $\mathcal{O}({\alpha}^2L^2) {\mu^+\mu^-}$,
  (e) 100 $\times$ $\mathcal{O}({\alpha}^3L^3) {\gamma}$, 
  (f) 100 $\times$ $\mathcal{O}({\alpha}^3L^3) {e^+e^-}$,
  (g) 100 $\times$ $\mathcal{O}({\alpha}^3L^3) {\mu^+\mu^-}$
  and
  (h) 1000 $\times$ $\mathcal{O}({\alpha}^4L^4) {\gamma}$
  vs. c.m.s. energy.
}
\end{figure}

Figure~\ref{fig-delta-l1l2l3} presents the contributions 
of the orders $\mathcal{O}({\alpha}^nL^n), n=1-4$
to the QED ISR.
The largest effect corresponds to the first order $\mathcal{O}({\alpha}L)$ term which varies from about $-25\%$
at the c.m.s. energy $\sqrt{s}=220$ GeV to $+14\%$ at $\sqrt{s}=500$ GeV. This contribution is approximately 10 times larger than the second order contributions and 100 (1000) than the third (fourth) order terms.

\begin{figure}[ht]
\centering
\includegraphics[width=0.7\linewidth]{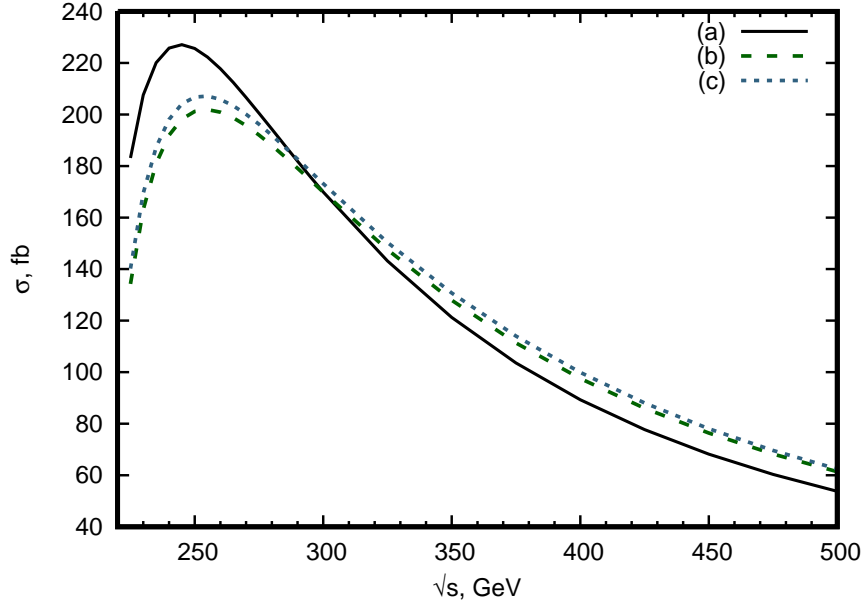}
\caption{\label{fig-sigma1}
  Cross-sections (in fb) vs. c.m.s. energy: 
  (a) the Born, (b) the one with $\mathcal{O}({\alpha})$ QED corrections,
  (c) the one with the complete one-loop EW contributions.
}
\end{figure}

Figure~\ref{fig-sigma1} illustrates the behaviour of the cross-sections with respect to the c.m.s. energy. It is seen that at the peak in the threshold region, the one-loop QED corrections change not only the height of the peak but also its form and position.

\begin{figure}[ht]
\centering
\includegraphics[width=0.7\linewidth]{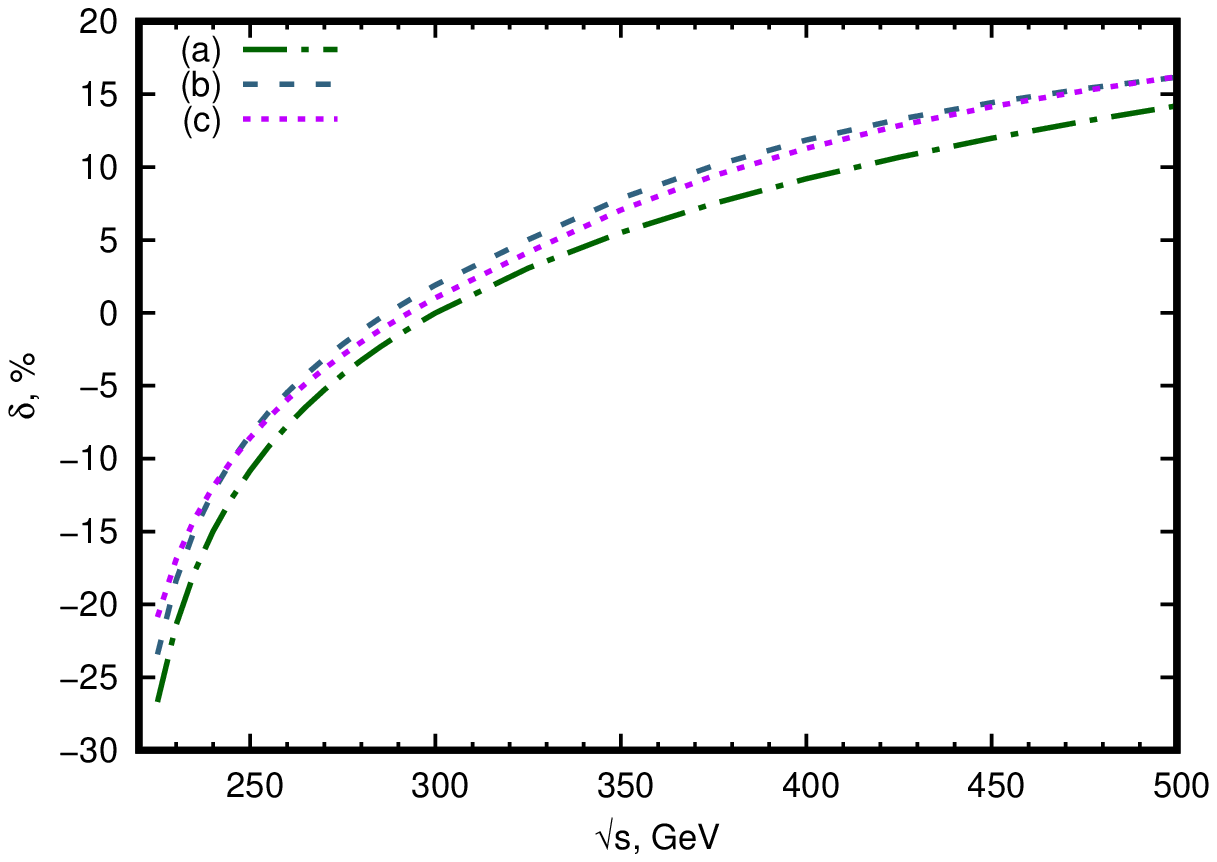}
\caption{\label{delta-qednlols}
  Relative corrections (in \%) (a) for the QED $\mathcal{O}({\alpha})$,
  (b) for the complete one-loop 
  and (c) for the sum of (b) and
  $\sum\limits_{n=2}^{4}\mathcal{O}(\alpha^nL^n)$ ISR
  contributions vs. c.m.s. energy.
}
\end{figure}

Figure~\ref{delta-qednlols} complements 
Fig.~\ref{fig-sigma1}. It shows the size of the relative RC in different approximations. One can see that the HO ISR LLA contributions 
provide a small but visible shift (the difference between lines (b) and (c)) from the complete one-loop EW correction. Moreover, this shift changes its sign.

In Tables~\ref{isrllatabshort240} and~\ref{isrllatabshort250} we show the 
ISR corrections of different order of ${\cal O}(\alpha^nL^n), n=2-4$ in the LLA approximation for
the c.m.s. energies $\sqrt{s}=240$~{GeV} and $250$~{GeV}
in the $\alpha(0)$ EW scheme.

{\small
\begin{table}[!ht]
\begin{center}
\begin{tabular}{|r|r|r|r|r|r|r|r|r|}
\hline
&
$\mathcal{O}(\alpha^2L^2)$ & 
$\mathcal{O}(\alpha^2L^2)$ &
$\mathcal{O}(\alpha^2L^2)$ &
$\mathcal{O}(\alpha^3L^3)$ &
$\mathcal{O}(\alpha^3L^3)$ &
$\mathcal{O}(\alpha^3L^3)$ &
$\mathcal{O}(\alpha^4L^4)$ &
$\sum\limits_{n=2}^{4}\mathcal{O}(\alpha^nL^n)$\\
&
$\gamma$ & 
$e^+e^-$ &
$\mu^+\mu^-$ &
$\gamma$ &
$e^+e^-$ &
$\mu^+\mu^-$ &
$\gamma$ &
\\
\hline
$\delta\sigma_{\text{\tiny{LLA}}}$, fb & $1.128(1)$ & $-0.368(1)$ & $-0.218(1)$ & $0.176(1)$ & $0.019(1)$ & $0.011(1)$ & $-0.023(1)$ & $0.727(1)$\\
%$\delta\bar{\sigma}_{\text{\tiny{LLA}}}$, fb &&&&&&&&$0.814(1)$\\
%$\delta\underline{\sigma}_{\text{\tiny{LLA}}}$, fb &&&&&&&&$0.646(1)$\\
\hline
$\delta_{\text{\tiny{LLA}}}$, \% & $0.500(1)$ & $-0.163(1)$ & $-0.097(1)$ & $0.078(1)$ & $0.008(1)$ & $0.005(1)$ & $-0.010(1)$ & $0.322(1)$\\
%$\bar{\delta}_{\text{\tiny{LLA}}}$, \% &&&&&&&&$0.361(1)$\\
%$\underline{\delta}_{\text{\tiny{LLA}}}$, \% &&&&&&&&$0.286(1)$\\
\hline
\end{tabular}
\caption{ISR corrections in the LLA approximation for the $e^+e^- \to ZH$ process at $\sqrt{s} = 240$ GeV. No cuts are imposed. Here $\delta_{\text{\tiny{ISR LLA}}} \equiv \delta\sigma_{\text{\tiny{ISR LLA}}}/\sigma_{0}$. The Born cross section is $\sigma_0 = 225.74(1)$ fb.}
\label{isrllatabshort240}
\end{center}
\end{table}
}

{\small
\begin{table}[!ht]
\begin{center}
\begin{tabular}{|r|r|r|r|r|r|r|r|r|}
\hline
&
$\mathcal{O}(\alpha^2L^2)$ & 
$\mathcal{O}(\alpha^2L^2)$ &
$\mathcal{O}(\alpha^2L^2)$ &
$\mathcal{O}(\alpha^3L^3)$ &
$\mathcal{O}(\alpha^3L^3)$ &
$\mathcal{O}(\alpha^3L^3)$ &
$\mathcal{O}(\alpha^4L^4)$ &
$\sum\limits_{n=2}^{4}\mathcal{O}(\alpha^nL^n)$\\
&
$\gamma$ & 
$e^+e^-$ &
$\mu^+\mu^-$ &
$\gamma$ &
$e^+e^-$ &
$\mu^+\mu^-$ &
$\gamma$ &
\\
\hline
$\delta\sigma_{\text{\tiny{LLA}}}$, fb 
& $-0.223(1)$ & $-0.268(1)$ & $-0.159(1)$ & $0.211(1)$ & $-0.010(1)$ & $-0.006(1)$ & $-0.016(1)$ & $-0.468(1)$\\
%$\delta\bar{\sigma}_{\text{\tiny{LLA}}}$, fb &&&&&&&&$-0.516(1)$\\
%$\delta\underline{\sigma}_{\text{\tiny{LLA}}}$, fb &&&&&&&&$-0.423(1)$\\
\hline
$\delta_{\text{\tiny{LLA}}}$, \% 
& $-0.099(1)$ & $-0.119(1)$ & $-0.070(1)$ & $0.094(1)$ & $-0.004(1)$ & $-0.003(1)$ & $-0.007(1)$ & $-0.207(1)$\\
%$\bar{\delta}_{\text{\tiny{LLA}}}$, \% &&&&&&&&$-0.228(1)$\\
%$\underline{\delta}_{\text{\tiny{LLA}}}$, \% &&&&&&&&$-0.187(1)$\\
\hline
\end{tabular}
\caption{ISR corrections in the LLA approximation for the $e^+e^- \to ZH$ process at $\sqrt{s} = 250$ GeV. 
No cuts are imposed. Here $\delta_{\text{\tiny{ISR LLA}}} \equiv \delta\sigma_{\text{\tiny{ISR LLA}}}/\sigma_{0}$.
The Born cross section is $\sigma_0 = 225.59(1)$ fb. 
%$\bar{\sigma}$ and $\bar{\delta}$ are calculated 
%at scale = $2\sqrt{s}$; $\underline{\sigma}$ and $\underline{\delta}$ are calculated at scale = $\sqrt{s}/2$.
}
\label{isrllatabshort250}
\end{center}
\end{table}
}

It is seen in Tables~\ref{isrllatabshort240} and \ref{isrllatabshort250} that
the corrections for %total value  
%${\delta\sigma^{\text{QED}}}$, i.e.
the sum of all considered orders of the ISR terms $\sum_{n=2}^4{\cal O}(\alpha^nL^n)$
are about 0.322\% for the c.m.s. energy $\sqrt{s}=240$~GeV and about -0.207\% for 
the c.m.s. energy $\sqrt{s}=250$~GeV. 
For the c.m.s. energy $\sqrt{s}=240$~GeV the most significant HO contribution is of course the 
photonic one of the order ${\cal O}(\alpha L)^2$. It composes about half a percent while 
from pairs we get about $-(0.1-0.2) \%$.
For the c.m.s. energy $\sqrt{s}=250$~GeV the dominant contributions of the second order are about
$-0.099\%$ for $\gamma$ and $-0.119\%$ for $e^+e^-$-pairs ($-0.070\%$ for $\mu^+\mu^-$-pairs), respectively.
When considering HO corrections, we see that it is certainly sufficient to take into account corrections 
up to the fourth order.

Variation of the factorization scale in the argument of the large logarithm can simulate 
the next-to-leading corrections, e.g., $\mathcal{O}(\alpha^2L)$. In the same manner as in
estimates of scale variation uncertainties in QCD, we apply factors 2 and 1/2 in the argument
of the large logarithm. 
This leads to the following values of the HO LLA corrections at $\sqrt{s}=240$~GeV:
$\delta_{\text{\tiny{LLA}}}(2\sqrt{s})=0.361(1)\%$ and 
$\delta_{\text{\tiny{LLA}}}(\sqrt{s}/2)=0.286(1)\%$, respectively. 
And for $\sqrt{s}=250$~GeV we get
$\delta_{\text{\tiny{LLA}}}(2\sqrt{s})=-0.228(1)\%$ and 
$\delta_{\text{\tiny{LLA}}}(\sqrt{s}/2)=-0.187(1)\%$. 

\begin{table}[!ht]
\begin{center}
\begin{tabular}{|r|r|r|r|}
\hline
EW scheme & $\alpha(0)$ & $G_\mu$ & $\alpha(M_Z^2)$\\
\hline
$\sigma_{\text{\tiny{Born}}}$, fb & $225.74(1)$ & $240.43(1)$ & $254.65$(1)\\
\hline
$\sigma_{\text{\tiny{Born+PW}}}$, fb & $231.88(1)$ & $233.25(1)$ & $231.80(1)$\\
\hline
$\delta\sigma_{\text{\tiny{PW}}}$, fb & $6.15(1)$ & $-7.18(1)$ &$-22.85(1)$ \\
\hline
$\delta_{\text{\tiny{PW}}}$, \% & $2.72(1)$ & $-2.99(1)$ & $-8.97(1)$\\
\hline
\end{tabular}
\caption{The Born and pure weak corrections in different EW schemes at the c.m.s. energy $\sqrt{s} = 240$ GeV.}
\label{Table:delta_240-sch}
\end{center}
\end{table}

\begin{table}[!ht]
\begin{center}
\begin{tabular}{|r|r|r|r|}
\hline
EW scheme & $\alpha(0)$ & $G_\mu$ & $\alpha(M_{\rm{Z}}^2)$\\
\hline
$\sigma_{\text{\tiny{Born}}}$, fb & $225.59(1)$ & $240.27(1)$ & $254.49(1)$\\
\hline
$\sigma_{\text{\tiny{Born+PW}}}$, fb & $231.17(1)$ & $232.49(1)$ & $231.01(1)$\\
\hline
$\delta\sigma_{\text{\tiny{PW}}}$, fb & $5.58(1)$ & $-7.78(1)$ & $-23.48(1)$\\
\hline
$\delta_{\text{\tiny{PW}}}$, \% & $2.47(1)$ & $-3.24(1)$ & $-9.22(1)$\\
\hline
\end{tabular}
\caption{The Born and pure weak corrections in different EW schemes at the c.m.s. energy $\sqrt{s} = 250$ GeV.}
\label{Table:delta_250-sch}
\end{center}
\end{table}

In Tables~\ref{Table:delta_240-sch}  and~\ref{Table:delta_250-sch},
the results of the Born cross sections,
the sum of the Born and pure weak (PW) contributions
as well as the relative corrections $\delta$ (\%)
for the c.m.s. energies $\sqrt{s}=240$ and $250$~GeV
in the $\alpha(0)$, $G_\mu$ and $\alpha(M_Z^2)$ EW schemes are presented.
The $\alpha(M_{\rm{Z}}^2)$ = 1/129.02 value was used in the calculations.
%The relative correction $\delta$ (in \%) is defined %as
%\bqa
%\delta = \frac{\sigma^{\text{Born+PW}}}{\sigma^{\text{Born}}} -1.
%\eqa
As it is seen the corrections in the
$\alpha(0)$ scheme are positive and equal to
2.72\% for the c.m.s. energy $\sqrt{s}=240$~GeV and
2.47\% for $\sqrt{s}=250$~GeV.
The calculations in the $G_{\mu}$ scheme reduce RC
to about 5-6 \%, they become negative and equal
to -2.99\% for the c.m.s. energy $\sqrt{s}=240$~GeV
and -3.24\% for the c.m.s energy $\sqrt{s}=250$~GeV.
In the case of the $\alpha(M_Z^2)$ scheme, RCs
get even more negative and achieve the
value -8.97\% and -9.22\%
for the c.m.s. energies $\sqrt{s}=240$~GeV and $\sqrt{s}=250$~GeV,
respectively. These results show that there is no most suitable EW scheme
of calculations for minimizing the value of the pure weak corrections 
for the $e^+ e^- \to ZH$ reaction. However, the sensitivity
to the choice of input EW scheme is reduced for the Born+PW 
cross-sections compared to the Born one. 
In~\cite{Sun:2016bel} and~\cite{Gong:2016jys},
the mixed QCD and EW NNLO corrections were considered and a further
reduction of the EW scheme dependence was observed.

%\begin{table}[!ht]
%\begin{center}
%\begin{tabular}{|r|r|r|r|r|}
%\hline
%&\multicolumn{4}{c|}{$\sum\limits_{n=1}^{N}\mathcal{O}(\alpha^nL^n)$}\\
%\hline
%$N$ & 1 & 2 & 3 & 4\\
%%&
%%$\mathcal{O}(\alpha^1L^1)$ & 
%%$\sum\limits_{n=1}^{2}\mathcal{O}(\alpha^nL^n)$ &
%%$\sum\limits_{n=1}^{3}\mathcal{O}(\alpha^nL^n)$ &
%%$\sum\limits_{n=1}^{4}\mathcal{O}(\alpha^nL^n)$ \\
%\hline
%\multicolumn{5}{|c|}{$\sqrt{s} = 240$ GeV}\\
%\hline
%$\sigma_{\text{\tiny{LLA}}}$, fb 
%& $190.467(1)$ & $191.595(1)$ & $191.771(1)$ & $191.748(1)$ \\
%\hline
%$\sigma_{\text{\tiny{Exp Add}}}$, fb 
%& $191.250(1)$ & $192.769(1)$ & $191.734(1)$ &  \\
%\hline
%$\sigma_{\text{\tiny{Exp Mul}}}$, fb 
%& $191.656(1)$ & $191.735(1)$ & $191.737(1)$ &  \\
%\hline
%\multicolumn{5}{|c|}{$\sqrt{s} = 250$ GeV}\\
%\hline
%$\sigma_{\text{\tiny{LLA}}}$, fb 
%& $199.733(1)$ & $199.511(1)$ & $199.722(1)$ & $199.706(1)$ \\
%\hline
%$\sigma_{\text{\tiny{Exp Add}}}$, fb 
%& $199.100(1)$ & $199.755(1)$ & $199.714(1)$ &  \\
%$\sigma_{\text{\tiny{Exp Mul}}}$, fb 
%& $199.601(1)$ & $199.716(1)$ & $199.719(1)$ &  \\
%\hline
%\end{tabular}
%\caption{Comparison between results with order-by-order and exponentiated structure functions. Only pure photonic corrections are taken %in account. Here $\sigma_{\text{\tiny{Exp Add}}}$ calculated with
%the electron structure functions taken in the additive exponentiated form \cite{Cacciari:1992pz} and $\sigma_{\text{\tiny{Exp Mul}}}$ in %the multiplicative exponentiated form \cite{Przybycien:1992qe}.
%}
%\end{center}
%\end{table}

\begin{table}[!ht]
\begin{center}
\begin{tabular}{|r|r|r|r|r|}
\hline
&\multicolumn{4}{c|}{$\sum\limits_{n=1}^{N}\mathcal{O}(\alpha^nL^n)$}\\
\hline
$N$ & 1 & 2 & 3 & 4\\
\hline
\multicolumn{5}{|c|}{$\sqrt{s} = 240$ GeV}\\
\hline
$R_{\text{\tiny{LLA}}}$ 
& $0.9934$ & $0.9993$ & $1.0002$ & $1.0001$ \\
\hline
$R_{\text{\tiny{Exp Add}}}$ 
& $0.9975$ & $1.0054$ & $1.0000$ &  \\
\hline
$R_{\text{\tiny{Exp Mul}}}$ 
& $0.9996$ & $1.0000$ & $1.0000$ &  \\
\hline
\multicolumn{5}{|c|}{$\sqrt{s} = 250$ GeV}\\
\hline
$R_{\text{\tiny{LLA}}}$ 
& $1.0001$ & $0.9990$ & $1.0000$ & $0.9999$ \\
\hline
$R_{\text{\tiny{Exp Add}}}$ 
& $0.9969$ & $1.0002$ & $1.0000$ &  \\
\hline
$R_{\text{\tiny{Exp Mul}}}$ 
& $0.9994$ & $1.0000$ & $1.0000$ &  \\
\hline
\end{tabular}
\caption{Comparison between results with order-by-order and exponentiated structure functions. Only pure photonic corrections are taken in account. Here
$R_{i} = \sigma_{i}/\sigma^{(3)}_{\text{{Exp Mul}}}$,
$i=(\text{{LLA}},\text{{Exp Add}},\text{{Exp Mul}})$,
$\sigma_{\text{{Exp Add}}}$ calculated with
the electron structure functions taken in the additive exponentiated form \cite{Cacciari:1992pz} and $\sigma_{\text{{Exp Mul}}}$ in the multiplicative exponentiated form \cite{Przybycien:1992qe}.
}
\label{table_exp}
\end{center}
\end{table}

In Table~\ref{table_exp}, we verified the difference between order-by-order and exponentiated ("additive" according to the prescription of Kuraev and Fadin \cite{Kuraev:1985hb} and "multiplicative" proposed by Jadach and Ward \cite{JADACH1990351})
realizations of the electron structure function. Results are shown up to $\mathcal{O}(\alpha^3L^3)$
finite terms for exponentiated forms and up to $\mathcal{O}(\alpha^4L^4)$ for order-by-order 
calculation. It can be seen that result using multiplicative exponentiated form converges faster.
But taking into account four orders in the order-by-order calculation is enough to reach 
the $10^{-4}$ accuracy.

\section{Conclusions \label{conclusion}}
%%%%%%%%%%%%%%%%%%%%%%%%%%%%%%%%%%%%%%%%%%

So we considered the contributions due to the QED initial state radiation 
(photons and pairs) to the higgsstrahlung process. Their impact has
been analyzed order by order. The complete one-loop electroweak one-loop
corrections were presented. Higher-order ISR QED contributions
were calculated within the leading logarithmic approximation.
The known expressions for contributions of the collinear electron structure function  of 
the orders ${\cal O}(\alpha^n L^n), n=2-4$ for photons and pairs
were used. These corrections are known to be very important 
in the case of resonances, e.g., at the $Z$-boson peak studied at LEP. 
We would like to emphasize that higher-order QED ISR corrections can be 
large not only at resonances but also near the reaction thresholds.
Note that the cross section of this process has a peak at the threshold.
%One can see that ${\cal O}(\alpha^4 L^4)$ ISR contributions can become relevant 
%if the level of accuracy of the order $\sim 10^{-5}$ is required. 

By looking at the magnitude of the complete one-loop electroweak
and higher-order LLA QED corrections, we can estimate the theoretical
uncertainty and define what other contributions should be taken into
account. Namely, a safe estimate of the theoretical uncertainty in 
EW and LLA RC can be derived by variation of the EW scheme and
the factorization scale, respectively. 
One can see that to meet high precision of future experiments,
we need to go beyond the approximation explored here. At least
the next-to-leading QED ISR logarithmic corrections should also be
taken into account. One needs to improve the uncertainty
in pure weak contributions. That can be done by taking into account 
higher-order EW and mixed QCD and EW effects in the $Z$ boson propagator 
and vertices. Note also that corrections for the whole processes with different
decay modes of $Z$ and Higgs bosons should be evaluated.

For the one permille precision tag relevant for future studies of 
the higgsstrahlung process, we see that there is 
a good agreement between the order-by-order results
and the known exponentiated QED LLA corrections~\cite{Cacciari:1992pz,Przybycien:1992qe}.
So either approach can be used. Presumably, the exponentiated one
is more suitable for Monte Carlo simulations, while the order-by-order
one can be used for benchmarks and cross-checks.

The numerical results presented here were obtained by means of the Monte Carlo generator 
ReneSANCe~\cite{Sadykov:2020any} and MCSANCee integrator~\cite{MCSANCee:2021} 
which allow evaluation of arbitrary differential cross sections.
These computer codes can be downloaded from the SANC project homepage~\url{http://sanc.jinr.ru} 
and the ReneSANCe HEPForge page~\url{https://renesance.hepforge.org}.

%%%%%%%%%%%%%%%%%%%%%%%%%%%%%%%%%%%%%%%%%%
\subsection*{Acknowledgments}
This research was funded by RFBR grant 20-02-00441.
The authors are grateful to Ya.~Dydyshka for fruitful discussions.

%%%%%%%%%%%%%%%%%%%%%%%%%%%%%%%%%%%%%%%%%%
\bibliography{ZH_ISR_QED_arxiv}

\end{document}